\documentclass[12pt,a4paper]{article}
\usepackage{latexsym,amsmath,amssymb}
\date{}
\numberwithin{equation}{section}

\begin{document}
\title{Causal Perturbation Theory for the Supersymmetric Wess-Zumino model}
\author{Florin Constantinescu\\ Fachbereich Mathematik \\ Johann Wolfgang Goethe-Universit\"at Frankfurt\\ Robert-Mayer-Strasse 10\\ D 60054
Frankfurt am Main, Germany }
\maketitle

\begin{abstract}
We illustrate the causal perturbation and causal renormalization method
(the Epstein-Glaser method) for the case of the supersymmetric
Wess-Zumino model. Our study is based on the Hilbert space structure of the
$N=1$ superspace.
\end{abstract}

\section{Introduction}

The causal perturbation theory of Epstein and Glaser \cite{EG} is a
renormalization method for the operator valued time-ordered products in
quantum fields theory. This elegant approach has been applied by its
inventors to the $\varphi^4 $-model in four dimensions and proved to be equivalent with the difficult BPHZ renormalization method. The main input of the method is causality in local quantum field theory. It is the only perturbative approach in which renormalization is explored
rigorously at the operator level and in which the unitarity of the scattering matrix operator is proved (in the sense of formal power series). The mathematical tools of the method are Hilbert space and distribution theory. The natural question which arises is whether the method can be extended to supersymmetric quantum field theory. This is in our opinion an interesting question taking into account that in supersymmetry results on canonical quantization methods and other well known methods in quantum field theory, including mathematical rigorous developements, are rather rare and the vigorous work is concentrated on path integrals. \\
In this paper we give a short study of the supersymmetric Wess-Zumino model by the Epstein-Glaser method. This study is based on our previous findings on positive definite Hilbert space and distribution theory in the supersymmetric framework \cite{FC}, in particular on what we called the Hilbert-Krein structure of $N=1$ supersymmetry.
From the technical point of view we use the notion of scaling order of distributions and the so called method of extension instead of cutting operator-valued distributions like in \cite{EG}. The extension method was used already a number of times \cite{BF,F}. \\
The result we obtain is renormalizability, as it should be, and as such
it is not new. What counts is the proof that this operaror method works
supersymmetrically and is susceptible of extensions. We use the
notations and conventions of \cite{WB} with the only difference that the
Pauli $\sigma_0 $ is one instead if minus one in \cite{WB} (our notations coincide with that of \cite{S} and \cite{FC}).

\section{Standard Hilbert-Krein superspace and superdistributions}

We write the most general (test) function $X=X(z)=X(x,\theta ,\bar \theta )$ as

\begin{gather}\nonumber
X(x,\theta ,\bar \theta )=f(x)+\theta \varphi (x) +\bar \theta \bar \chi (x) +\theta ^2m(x)+\bar \theta^2n(x)+ \\ 
\theta \sigma^l\bar \theta v_l(x)+\theta^2\bar \theta \bar \lambda(x)+\bar \theta^2\theta \psi (x)+ \bar \theta^2 \theta^2d(x)
\end{gather}
where the coefficients are functions from the Schwartz space $S$ of smooth rapidly decreasing functions of $x $ in Minkowski space with signature (-1,1,1,1). Consider the tempered measure $d\rho (p)$ with support concentrated in the open backward light cone $V^- $ and its Fourier transform

\begin{gather}
D^+ (x)=\frac{1}{(2\pi )^3}\int e^{ipx}d\rho (p)
\end{gather}
Define the supersymmetric kernels

\begin{equation}
k(x,\theta ,\bar \theta )=k(z)=\delta^2 (\theta )\delta^2(\bar \theta
)\delta^4 (x)
\end{equation}
and

\begin{equation}
K_0(x,\theta ,\bar \theta )=K_0 (z)=\delta^2 (\theta )\delta^2(\bar
\theta )D^+ (x)
\end{equation}
where $K_0 $ is obtained from $k$ by replacing the delta-function $\delta^4 (x)$ by (2.2).
Define the inner products of supersymmetric $X$ and $Y$ by 

\begin{gather}
<X,Y>=\int dz_1 ^8 dz_2 ^8\bar X(z_1 )k(z_1 -z_2 )Y(z_2 )
\end{gather}
and

\begin{gather}
<X,Y>=\int dz_1 ^8 dz_2 ^8\bar X(z_1 )K_0 (z_1 -z_2 )Y(z_2 )
\end{gather}
where $d^8 z=d^4 xd^2 \theta d^2 \bar \theta $. For general supersymmetric functions they are formal generalizations of $L^2 $- scalar product on functions.  Unfortunately both of them are indefinite. Let us transform the inner product (2.6) as follows. First introduce for (2.6) the short hand notation

\begin{gather}
<X,Y>=\int \bar XK_0 Y
\end{gather}
We write

\begin{gather}\nonumber
\int \bar XK_0 Y=\int \bar X(P_c +P_a +P_T )K_0 Y=\int \bar X(P_c^2
+P_a^2 +P_T^2 )K_0 Y=\\
=\int (\bar X_a K_0Y_a +\bar X_c K_0Y_c +\bar X_T K_0Y_T )
\end{gather}
Here 
\[P_c =\frac{1}{16\square } \bar D^2 D^2,P_a =\frac{1}{16\square }D^2 \bar D^2,P_T =-\frac{1}{8\square } D^\alpha \bar D^2 D_\alpha \]
are the formal projection operators on the chiral, antichiral and
transversal sectors in superspace respectively and $ X_c =P_c X,X_a =P_a
X,X_T=P_T X $. Note that the support restriction on the measure $d\rho $ takes care of the d'Alembertian in the denominators in (2.8). With some mild restrictions of test functions we can circumvent the support restriction on the measure (allowing for instance "zero-mass" \cite{FC}) but this will not interest us here. \\
In \cite{FC} we have identified the (unexpected) origin of indefiniteness of the inner product (2.8): it is the plus sign in the front of $P_T $ in $ P_c +P_a +P_T =1$. Indeed the right positive definite (in fact at this stage only non-negative) scalar product in superspace extending the usual $L^2 $-product is

\begin{gather} \nonumber
(X,Y)=\int \bar X(P_c +P_a -P_T )K_0 Y=\int \bar X(P_c^2 +P_a^2 -P_T^2
)K_0 Y= \\
=\int (\bar X_a K_0Y_a +\bar X_c K_0Y_c -\bar X_T K_0Y_T )
\end{gather}
Now let us eliminate the zero vectors in (2.9). Certainly we have to
restrict the coefficient functions to the nontrivial support of the
measure $d\rho (p)$. It turns out that this  restriction is sufficient
to make the scalar product (2.9) strictly positive. This can be proved
either in an abstract way or by brute force computations \cite{FC}.\\
By a mild reformulation these considerations make clear that in $N=1$
supersymmetry we have a natural Krein structure which generates, via the
usual construction, what we have called the standard supersymmetric
Hilbert-Krein space. As long as we are concerned with supersymmetric
free fields or with perturbation theory taking place in the Hilbert
space of the free field (like for instance causal perturbation theory of
Epstein and Glaser at least in the abelian case) we are forced to use this structure. Certainly if we study only fields in the chiral/antichiral sector (like for instance the Wess-Zumino model in this paper) then the transversal contribution can be entirely neglected.\\
After we have clarified the Hilbert space matter of $N=1$ supersymmetries we stick to the superdistributions. Superdistributions have been considered several times in the literature and many results have been obtaind by using functional-analytic methods (for a review which contains also recent references see \cite{FP}). Nevertheless we prefer here to use our findings about the supersymmetric Hilbert space in order to advance a Gelfand tripple procedure. Indeed our scalar product (2.9) can be modified by introducting some Sobolev type weights in such a way to obtain a countable number of Hilbert norms. They induce a locally convex topology on the space os supersymmetric test functions. The technicalities are simple and by duality we obtain a supersymmetric extension of the standard Gelfand tripple $S\subset L^2 \subset S' $ and in particular define the space of (tempered) superdistributions $S'$ and of superdistributions $\cal{D'} $. Many useful properties stay valid in this frame (like for instance nuclearity). Certainly a definitions of supersymmetric distributions can be imagined in terms of the components of the test superfunctions; both constructions are equivalent.

\section{The free chiral/antichiral superymmetric field (free
  Wess-Zumino field)}

In this section we consider the free chiral/antichiral supersymmetric field which formally coincides with the one defined through the free supersymmetric Wess-Zumino Lagrangian \cite{WB,S}:

\begin{equation}
L=\int d^4 xd^2 \theta d^2 \bar \theta \bar \phi \phi +\frac{1}{2}m(\int d^4 xd^2 \theta  \phi^2 +\int d^4 xd^2 \bar \theta \bar \phi^2 )
\end{equation}
There are two possibilities to include full Grassmann integration in the mass term, either writig $L$ as

\begin{equation}
L=\int d^4 xd^2 \theta d^2 \bar \theta \bar [\bar \phi \phi
+\frac{1}{2}m (\phi^2 \delta^2 (\bar \theta ) + \bar \phi^2 \delta^2 (\theta ))]
\end{equation}
or as

\begin{equation}
L=\int d^4 xd^2 \theta d^2 \bar \theta [\bar \phi \phi -\frac{1}{8}m (\frac{D^2}{\square }\phi^2 + \frac{\bar D^2 }{\square }\bar \phi^2 )]
\end{equation}
We will realize this field as an opertator-valued superdistribution in the symmetric Fock space defined over a Hilbert space  of supersymmetric functions. This Hilbert space can be constructed with the help of the standard Hilbert-Krein structure introduced in Section 2. This was already donne in \cite{FC}. For the convenience of the reader we give here the construction in a slightly modified way starting from scratch. The approach starts by considering the following matrix-operator in superspace

\begin{gather}
M=\begin{pmatrix}\square P_c & \frac{m}{4}\bar D^2 &0 \\ \frac{m}{4}D^2 & \square P_a &0\\0&0&-\square P_T \end{pmatrix}=\begin{pmatrix}\frac{1}{16}\bar D^2D^2 & \frac{m}{4}\bar D^2 &0\\ \frac{m}{4}D^2 &\frac{1}{16}D^2 \bar D^2 &0\\0&0&\frac{1}{8}D^{\alpha }\bar D^2 D_{\alpha } \end{pmatrix}
\end{gather}
which contains the 2x2 chiral/antichiral block and the transversal
contribution in the right corner on the bottom. 
Note the same unexpected sign in front of $P_T$ which we have already encountered in (2.9). Without it the diagonal would sum up to $\square (P_c +P_a +P_T) =\square $. \\
Now suppose that the measure $d\rho (p)$ is given by $d\rho (p)=\theta(-p_0)\delta (p^2 -m^2 ), $ supported by the hyperboloid $p^2 =m^2 $ in the  backward light cone and consider as above

\begin{equation}\nonumber
D^+ (x)=\frac{1}{(2\pi )^3}\int e^{ipx}d\rho (p)=\frac{1}{(2\pi )^3}\int e^{ipx}\theta(-p_0)\delta (p^2 -m^2 )
\end{equation}
and 

\begin{equation}\nonumber
K_0(x,\theta ,\bar \theta )=\delta^2 (\theta )\delta^2 (\bar \theta )D^+ (x)
\end{equation}
Note that in this case $D^+ $ is up to a factor of $-i$ the Pauli-Jordan function (the measure $d\rho $ is suported in the backward instead of forward light cone because of the "most positive" Minkowski metric).
Let 
\begin{gather}
X=\begin{pmatrix}X_1 \\X_2 \\X_3 \end{pmatrix},Y=\begin{pmatrix}Y_1 \\Y_2 \\Y_3 \end{pmatrix}
\end{gather}
be triples of arbitrary supersymmetric functions and let $K_0 (z)=K_0 (z)I_3 $ where $I_3$ is the 3x3 identity matrix. Then we introduce the supersymmetric kernel

\begin{gather}
MK_0 (z)=\begin{pmatrix}\frac{1}{16}\bar D^2D^2 & \frac{m}{4}\bar D^2 &0\\ \frac{m}{4}D^2 &\frac{1}{16}D^2 \bar D^2 &0\\0&0&\frac{1}{8}D^{\alpha }\bar D^2 D_{\alpha } \end{pmatrix}K_0 (z)
\end{gather}
and the scalar product

\begin{gather}
(X,Y)=\int d^8 z_1 d^8 z_2 \bar X^T (z_1 )M K_0 (z_1-z_2 )Y(z_2 )
\end{gather}
Note that in (3.7) the covariant derivatives $D$ and $\bar D$ are applied on the first variable $z_1 $. They can be moved to the second variable $z_2 $ taking into account the relations

\begin{gather}
D_1^2 K_0 (z_1 -z_2 )=D_2^2 K_0 (z_1 -z_2 ) \\
D_1^2 \bar D_1^2 K_0 (z_1 -z_2 )=D_2^2 \bar D_2^2 K_0 (z_1 -z_2 ) \\
D_1^2 \bar D_1^2K_0 (z_1 -z_2 )=D_2^2 \bar D_2^2K_0 (z_1 -z_2 ) \\
D_{1\alpha }\bar D^2_1 D_{1\alpha }K_0 (z_1 -z_2 )=D_{2\alpha }\bar D^2_2 D_{2\alpha }K_0 (z_1 -z_2 )
\end{gather}
and their conjugates. Using these relations we can in fact distribute at will the derivatives to the first, the second or both variables. This fact is related to the possibility of introducing the equations of motions to be given below in a consistent way. The scalar product is non-negative but it can have zero vectors \cite{FC} stemming only from the chiral/antichiral sector. On-shell restriction of the coefficient functions is no longer sufficient but the zero vectors can be eliminated for instance by usual factorization followed by completion. For the particular case we are faced with in this paper, we prefer a much simple constructive way which will be given later on in this Section.\\
With the help of this scalar product defined on test functions with coefficients in the Schwartz space $S$ of rapidly decreasing functions we construct a Hilbert space and the associated (symmetric) Fock space with the vacuum $\Omega $ being the supersymmetric function one.
First because we will study a model only in the chiral/antichiral sector (the Wess-Zumono model) we cut away the transversal sector staying with the scalar product (3.7) where now the kernel in the supersymmetric integral is given by

\begin{gather}
MK_0 (z)=\begin{pmatrix}\frac{1}{16}\bar D^2D^2 & \frac{m}{4}\bar D^2 \\ \frac{m}{4}D^2 &\frac{1}{16}D^2 \bar D^2 \end{pmatrix}K_0 (z)
\end{gather}
and restrict ourself to the easier massive case $m>0$. The massless case is a little more involved and is studied in \cite{FC}. Note in passing that the operator matrix $M$ can be read up from physical work \cite{WB}. As before the derivatives act on the first variable of the kernel. 
We define the free chiral/antichiral quantum field 

\begin{equation}
\Phi (x,\theta ,\bar \theta )=\begin{pmatrix}\phi (x,\theta ,\bar \theta )\\ \bar \phi (x,\theta ,\bar \theta )\end{pmatrix}
\end{equation}
or in the smeared form
\begin{equation}
\Phi (F)=\begin{pmatrix}\phi (f_1)\\ \bar \phi (f_2)\end{pmatrix}
\end{equation}
where $F=\begin{pmatrix}f_1 \\ f_2 \end{pmatrix}$ with $f_1,f_2 $ supersymmetric, through its two point function

\begin{gather}\nonumber
\omega (z_1 ,z_2 )=\begin{pmatrix}(\Omega ,\phi (z_1 ) \bar \phi (z_2 )\Omega )& (\Omega ,\phi (z_1 ) \phi (z_2 )\Omega ) \\ (\Omega ,\bar \phi (z_1 ) \bar \phi (z_2 )\Omega )& (\Omega ,\bar \phi (z_1 ) \phi (z_2 )\Omega )\end{pmatrix}= \\
=\begin{pmatrix}\frac{1}{16}\bar D^2D^2 & \frac{m}{4}\bar D^2 \\ \frac{m}{4}D^2 &\frac{1}{16}D^2 \bar D^2 \end{pmatrix}K_0 (z_1 -z_2 )
\end{gather}
It is clear that we have here rigorously defined the free field given by
the free Wess-Zumino Lagrangian (3.3) \cite{WB}.   
Certainly we can restrict the test functions without loose of information to the chiral/antichiral sector. Consistency implies that the chiral component $\phi $ in $\Phi $ smeared with a chiral function vanishes and the similar statement in the antichiral case.
We use also the self-explanatory notation in the smeared form
\begin{equation}
(F,\omega G)=\begin{pmatrix}(f_1 ,\omega_{cc}g_1 ) & (f_1 ,\omega_{ca}g_2 ) \\ (f_2 ,\omega_{ac}g_1 ) & (f_2 ,\omega_{aa}g_2 ) \end{pmatrix}
\end{equation}
where $F$ and $G$ are of the form above.
Note thet $\omega (z)$ is not translation invariant in $z$ but fortunately it is in the space-time variables.\\
At this stage, after rigorously defining the general free chiral/antichiral field (i.e. off mass shell and with auxiliary fields), we want to particularize it for the purposes of this paper. Indeed the Green functions in the Epstein-Glaser construction are on shell and the auxiliary fields carry in this framework no special information. We will reduce the problem as follows. \\
Let us consider only test functions of the form
\begin{equation}
F=\begin{pmatrix}f \\\bar f \end{pmatrix}
\end{equation}
where $\bar f $ is the conjugate of $f$. For $f$ we introduce equations of motion as follows

\begin{equation}
mf =\frac{1}{4}\bar D^2 \bar f
\end{equation}
equivalent to

\begin{equation}
m\bar f =\frac{1}{4}D^2 f
\end{equation}
It can be proven that imposing the equations of motion the scalar product defined in (3.7) is stricty positive definite without any further factorization. On the other hand these considerations show that we can reduce the problem to a single non-neutral scalar field $\phi $ (the chiral field) with a strict positive definite two point function 
\begin{equation}
\omega_{cc} (z_1 ,z_2 )=\frac{1}{16}\bar D^2 D^2 K_0(z_1 -z_2 )
\end{equation}
in the corresponding restricted Fock space with vacuum being again the constant function one. In order to avoid inflationary notations we will continue to write $\omega (z_1 ,z_2 )$ for $ \omega_{cc} (z_1 ,z_2 ) $.  The antichiral field $\bar \phi $ acts in this space as $\frac{1}{4m}D^2 \phi $.  We can work in either one of these representations. The second representation is sometime easier to handle. When working in this second representation we will use the notation $\Phi (z)$ to understand either $\phi (z)$ or $\bar \phi (z)$.\\
In order to saddle down for causal perturbatin theory \cite{EG} we have first to define Wick order and to prove the Wick theorem. The definition of Wick order is recursive as in the usual case : $ :1:=1, :\Phi (z):=\Phi (z) $ and

\begin{gather}
:\Phi (z)\prod_{i=1}^n \phi (z_i ):=\Phi (z):\prod_{i=1}^n \Phi (z_i ):-\sum_{k=1}^n \omega (z,z_k ):\prod_{i\in {1,\ldots ,n},i\neq k}\Phi (z_i ):
\end{gather}
Wick powers ,Wick monomials and Wick polynomials are produced from the inductive definition by letting some arguments coincide or, if a distribution-theoretic framework is prefered, by using improper test functions of the type
$\delta^4 (x_1 -x_2 )\delta^2 (\theta_1 -\theta_2 )\delta^2(\bar \theta_1 -\bar \theta_2 )$ (supersymmetric Dirac function).
We have to show that these definitions are equivalent to the usual definitions on component fields. This follows from the fact that our ad hoc definition of the two point function through the operator matrix $M$ (3.12) can be obtained in two ways: either by Gaussian integration in superspace (producing the propagators which afterward are turned into two point functions) or by a computation on components using the two point functions of the component fields. The fact that both methods produce the same result can be inferred from \cite{EG}. By the same argument the supersymmetric Wick theorem which follows from the definition (3.21) is equivalent with the Wick theorem on components. The easiest way to see this is to use an equivalent description of the Wick products by means of the generating function:

\begin{gather}
:e^{i\Phi (F)}:=e^{\frac{1}{2}\int \bar F \omega F}e^{i\Phi (F)}
\end{gather}
where in the exponent we have the supersymmetric integral with the two-point function $\omega $ as integral kernel. The Wick monomials can be constructed by multiplying this formula with itself a number of times. \\
After clarifying the situation with the (supersymmetric) Wick products and Wick theorem, it is easy to see that the "zero theorem" of Epstein and Glaser generalizes straightforward to the supersymmetric setting. This is an inocent looking theorem but central for this approach which deals with operator valued time ordered products as operator valued distributions. In our case it says that Wick ordered monomials multiplied by tempered superdistributions are operator-valued superdistridutions on an invariant domain which can de specified by algebraic vectors in Fock space of supersymmetric functions with regular coefficients of rapid decrease. This result enables to keep track of operator aspects in the process of induction to follow, in particular makes possible multiplication of field operators.\\ 
Before ending this section on the free chiral/antichiral field we describe also the interaction Lagrangian $\phi^k ,\bar \phi^k ,k=3 $ in the Wess-Zumino model. It is, up to the coupling constant \cite{WB}

\begin{gather}
L'=\int d^4 xd^2 \theta :\phi^k :+\int d^4 xd^2 \bar \theta :\bar \phi^k :=\int d^4 xd^2 \theta \phi^k +\int d^4 xd^2 \bar \theta \bar \phi^k 
\end{gather} 
On the r.h.s., in the last expression, we have left out the Wick ordering because it leaves invariant powers of pure chiral or antichiral fields. This follows from the particular form of the non-diagonal elements of the operator-matrix $M$. Indeed single powers of the form $D^2 $ or $\bar D^2 $ leave untouched one of the Grassmann delta-function in $K_0$ which vanishes for coinciding arguments.\\
Exactely as for the free Lagrangean at the begining of this section, there are two possibilities to insert the missing Grassmann integral on $\theta ,\bar \theta $, compensating it by delta functions or by the operators $\frac{D^2}{\square},\frac{\bar D^2}{\square}$. 

\section{The supersymmetric Epstein-Glaser method}

The renormalization method of Epstein und Glaser \cite{EG} is an
inductive method for defining the operator valued time ordered products
(Green functions) $T_n (x_1 ,x_2 ,\ldots ,x_n )$ as formal power series
by heavily using locality and causality. Being an operator method it
needs ad initio a Hilbert space in which the operators and the
subsequent operations on them are well defined. All variants of the
method have a first algebraic formal part which is mainly the induction
on $n$ and a second part which is of functional analytic nature, closely
following the induction, and is responsible for the well definiteness of
$T_n (x_1 ,x_2 ,\ldots ,x_n )$ as operator valued distributions. It is
this second part which decides about the renormalizability of the
Lagrangian model under study. The functional analytic part of the method
mainly counts the order of singularity (also called singular order) of
the inductively constructed $T_n (x_1 ,x_2 ,\ldots ,x_n )$. If the order
of singularity stays bounded with $n$ growing indefinitely, the theory
is said to be renormalizable. Certainly there are much more details in
this dificult construction, at least when we want to remove the
singularites by minimal number of substractions and  to make contact to
important physical aspects as for instance symmetries of the theory,
Lagrangian counterterms and last but not least if we want to relate it to other renormalization methods.\\ 
There are in principle two variants for conducting the induction on $n$ in $T_n (x_1 ,x_2 ,\ldots ,x_n )$; one involves cutting  distributions by multiplying them with a step function (this was done in the original paper of Epstein and Glaser) and a second variant which is concerned with  extensions of distributions in variables $x_i ,i=1,2,\ldots n)$ from outside to the diagonal $x_i =x_j ,i\neq j $ (see \cite{BF,F,P}) and uses for that the rough notion of order of singularily (called also singular order). When one wants to get only information about renormalizability by controlling the inductive singular order, the second variant is recommended. In particular this is the case for theories without translation invariance \cite{BF,F}.\\

In this paper we want to illustrate the applicability of the causal renormalization for supersymmetric models by restricting ourselves to the rigorous inductive study of the order of singularity of the time-ordered products in the case of (the massive) Wess-Zumino model. First let us remark that the algebraic part of the Epstein-Glaser induction, as presented for instance in \cite{BF,F} as well as in the elementary paper \cite{P}, suffers no alteration by going from the usual to the supersymmetric case. The real problems are of functional-analytic nature. As already said, we restrict in the frame of this method, to a rigorous counting of the singular order of the time-ordered products. In particular we will not attempt to construct time-ordered products which safisfy the requirement of supersymmetry (or even Lorentz covariance) nor will we discuss unitarity  (albeit our Hilbert-Krein construction is an invitation to study unitarity of the supersymmetric $S$-matris by operator valued formal series ; our aim is merely to illustrate the applicability of the method). \\
Our strategy (inspired from physical work; see for instance
\cite{G,W})is to define first a supersymmetric singular order for the
(operator valued) time ordered products $T_n (x_1 ,\theta _1 ,\bar
\theta _1 ;x_2 ,\theta _2 ,\bar \theta  _2 ;\ldots ;x_n ,\theta _n ,\bar
\theta _n )=T(z_1 ,z_2 ,\ldots ,z_n )$  which depends on the Grassmann
variables $\theta $ and $\bar \theta $ and to prove that in the process
of induction it stays bounded when $n$ goes to infinity. We could define this property as superrenormalizability. Now there is a class of superdistributions (called homogeneous in the sequel) for which the supersymmetric singular order of the superdistribution in question is simply related to the usual singular order of its components such that the integrals

\begin{gather} \nonumber
T_n (x_1 ,x_2 ,\ldots ,x_n )= \\
=\int d^8 z_1 d^8 z_2 \ldots d^8 z_n T_n (x_1 ,\theta_1 ,\bar \theta_1 ;x_2 ,\theta_2 ,\bar \theta_2 ;\ldots ;x_n ,\theta_n ,\bar \theta_n )
\end{gather}
considered as usual distributions have a singular order which also stay bounded with growing $n$. Luckily (see Section 6) the superdistributions which are generated in the induction process are homogeneous indeed (in fact this is evident, being the main point of the supersymmetric power counting in physics). By this usual renormalization follows. In the next section we will be concerned with the order of singularity in the supersymmetric context.

\section{Scaling degree and singular order of superdistributions}

A distribution $T\in D'(R^d )$ is said to have scaling degree $s$
\cite{BF,F,P} at $x=0$, if $s$ is the infimum of $s'\in R $ such that
$\lambda^{s'}T(\lambda x)\to 0$ for $ \lambda \to 0 $. The limit is
understood in the sense of distributions. We define the order of
singularity (or singular order) of $T$ to be the real number
$\omega=[s]-d$ where [s] gives the largest integer smaller or equal to
$s$. The $d$-substraction is a result of the space-time integration. As
it should be, with increasing positive $\omega $ the distribution $T$ is
more and more singular, whereas the distributional (i.e. singular)
nature totally disappear for $\omega <0$. The case $\omega =0$ is on the
border of "logarithmic type" singularity. Differentiation increases the
order of singularity whereas multiplication with the variable and
integration decrease it. The following two results are of special use:
first multiplication with regular functions does not increase the
singular order and second, tensor product of distributions results in
addition of singular order. The point is that such a rough object
suffices for the characterisation of the singularity strength of a
distribution not only in one but also in several variables! An important
result is that distributions in $d\ge 1$-variables with non-negative
order which belong to $D'(R^d \setminus 0) $ can be extended to belong to $D'(R^d )$ (i.e. extended from outside to the diagonal). There are plenty of such extensions involving arbitrary constants and the art of the matter in applications is to find a distinguished minimal one (i.e. one which is as less arbitrary as possible). In this paper we will not be concerned with minimality nor with so called normalization conditions which in applications further restrict the arbitrariness.\\

Now suppose that we are concerned with superdistributions. Instead of using super test functions for studying their singularities we prefer a hand-made definition. Beside scaling degree and singular order of their components as functions of the space-time variables, we attach a new (Grassmann) contribution to the scaling (and singularity) order of $-\frac{1}{2}$ for each single Grassmann variable $\theta_i $ and $\bar \theta_i ,i=1,2 $ in $\theta =\begin{pmatrix}\theta_1 \\ \theta_2 \end{pmatrix},\bar \theta =\begin{pmatrix}\bar \theta_1 \\ \bar \theta_2  \end{pmatrix} $\cite{G,W} and, taking into acount all variables, Grassmann or not, define by addition a total scaling degree. This is motivated by the fact that multiplication by Grassmann variables has a two-fold tendency to make things better (i.e. less singular), either because too many Grassmann factors produce a vanishing result or, if this is not the case because we do not have enough of them, the subsequent Grassmann integration picks up less singular coefficient functions. Accordingly Grassmann integration AND differentiation introduce a scaling correction of $\frac{1}{2}$ per variable (both of them diminish the number of Grassmann variables). In order to obtain the singular order from the scaling degree we have to substract one for each integrated space-time variable and to add one half for an integrated Grassmann variable.\\
The reader should have already noted that the statements above are non-ambiguous only if the superdistributions in question have a certain kind of homogeneity. In order to understand what we mean by that let us write down the explicit expression of a chiral function/distribution. It is \cite{WB,FC}

\begin{gather}\nonumber
X(x,\theta ,\bar \theta )=f(x)+\theta \varphi (x) +\bar \theta \bar \chi (x) +\theta ^2m(x)+\bar \theta^2n(x)+ \\ \nonumber
\theta \sigma^l\bar \theta v_l(x)+\theta^2\bar \theta \bar \lambda(x)+\bar \theta^2\theta \psi (x)+ \bar \theta^2 \theta^2d(x)
\end{gather}
with the following restrictions on the coefficients:

\begin{gather}\nonumber
\bar \chi =\psi =0,n=0,v_l =\partial_l (if)=i\partial_l f, \\ 
\bar \lambda =-\frac{i}{2}\partial_l \varphi \sigma^l ,d=\frac{1}{4}\square f
\end{gather}
It is clear that the derivatives in (5.1) are such that they compensate in a homogeneous way the scaling degree of the Grassmann factors. But the kernel $K_0$, which will be mainly involved in the computations to follow, is neither chiral nor antichiral. This makes no problem because it is itself homogenous (being a monomial on $\theta ,\bar \theta $) or it appears under the action of homogeneous square covariant derivatives pushing it to the chiral/antichiral sector. \\
Before passing to the supersymmetric extension problem let us remark that there is still a point which deserves special attention. The question is what happens with the supersymmetric singular order by multiplication of superdistributions. Suppose we multiply two superdistributions such that the multiplication of the coefficient distributions is tensorial but arbitrary in Grassmann variables. The orders of singularity for the coefficients get added but some Grassmann products may produce a zero result. Because the contribution of a Grassmann variable to the singular order is negative it follows that adding supersymmetric singular orders of factors may give only an upper bound for the singulat order of the product. Certainly if the factors are homogeneous the bound will be exact.\\
Now it is clear that superdistributions with coefficient distributions defined on $R^d \setminus 0 $ can be extended in a unique or non-unique way to superdistributions on $R^d $ by extending the corresponding coefficient distributions. Certainly this extension could be not economic because usually the superdistributions we are dealing with in applications have coefficients related through derivatives (like for instance the chiral and antichiral ones) and as such their extensions could be related. But at the prize of non-uniticity the simple remark above does the extension job and this is all we need in the next section.

\section{Renormalizability of the supersymmetric Wess-Zumino model by causal renormalization theory} 

In this section we will provide the rigorous counting of the singular order of the time-ordered products in the frame of the Epstein-Glaser induction. We want to remark that in this paper the counting proceeds on the supersymmetric time-order products $T_n(z_1 ,z_2 ,\ldots ,z_n )=T_n (x_1 ,\theta _1 ,\bar \theta _1 ;x_2 ,\theta _2 ,\bar \theta  _2 ;\ldots ;x_n ,\theta _n ,\bar \theta _n )$ which contain the Grassmann variables in the non-integrated form. Certainly, alternatively, we could attempt to set down an induction directly on the set of Grassmann integrated time-ordered products $T_n (x_1 ,x_2 ,\ldots ,x_n )=\int d^2 \theta_1 d^2 \bar \theta  \ldots d^2 \theta_n \bar \theta_n)T_n(z_1 ,z_2 ,\ldots ,z_n )$. At the first glance it seems that this alternative way is better suited because, as is well-known, Grassmann integration induces "miraculous" cancellations between component fields which could be of much help. Nevertheless it turn out that such a procedure (which more or less would be equivalent to Epstein-Glaser method on each component field separately) is cumbersome at the level of the induction itself. This is the reason we prefer the first method (we will see not too much of the "miraculous" cancellations and "renormalization theorem" at the stage of this paper but the method is likely to be improved). In this way we stay on the safety ground of the supersymmetric Hilbert space.\\
We use the notations and relations from \cite{BF,F}and the elementary presentation \cite{P} which we mainly follow. In particular we use for the time ordered products notations like

\begin{gather}
T_n (z_1 ,z_2 ,\ldots ,z_n )=T(L'(z_1)L'(z_2 )\ldots L'(z_n )) \\
T_{|I|}(z_i ,i\in I)=T(\prod _{i\in I}L'(z_i ))
\end{gather}
We perform the induction based on the following two hypothesis: \\

H1. Causality: let $N={1,2,\ldots ,n},I\subset N, I\cup I^c =N $ with
disjoint $I,I_c $ and introduce the short hand notation
$T_{|I|}(|I|)=T_{|I|}(z_i ,i\in I)$. We write $x\ge y$ iff $x$ is not in $\bar V^- (y)$ and use the notation $I\ge J$ with obvious meaning. Here $\bar V^- (y) $ and $\bar V^+ (y)$ are the closed backward and forward light cones respectively. Then for $I\ge I^c $ the T-products factorize:

\begin{gather}
T_n (N)=T_{|I|}(I)T_{|I^c |}(I^c ) 
\end{gather}
and \\

H2. Translation covariance: under a space-time translation the $T$-products transforms like

\begin{gather}\nonumber
U(a)T_n (x_1 ,\theta _1 ,\bar \theta _1 ;x_2 ,\theta _2 ,\bar \theta  _2 ;\ldots ;x_n ,\theta _n ,\bar \theta _n )U^{-1}(a)= \\
=T_n (x_1 +a,\theta _1 ,\bar \theta _1 ;x_2 +a,\theta _2 ,\bar \theta  _2 ;\ldots ;x_n +a,\theta _n ,\bar \theta _n )
\end{gather}
where $U(a)$ represents the translation by the vector $a$.\\
Now we are at the point of starting the induction. We set $T(0)=1,T_1 (z)=L'(z_1 )$. By the induction hypothesis we assume that for all $n'<n$ the time-ordered products $T_{n'}(z_1 ,z_2 \ldots ,z_{n'} )$ exists as operator valued superdistributions in Fock space and are of the form

\begin{gather}
T_{n'}(z_1 ,z_2 ,\ldots ,z_{n'} )=\sum_{l_i <k} t_{n'}^{l_1 ,\ldots ,l_{n'}}(z_1 ,z_2 ,\ldots ,z_{n'} ):\Phi^{k-l_1  }(z_1 )\ldots \Phi^{k-l_{n'}}(z_{n'}):
\end{gather}
with $t_{n' }$ being a scalar superdistribution on $R^{4n'}$. In the frame of induction we show first that $T_n (z_1 ,z_2 ,\ldots ,z_n ) $ is well defined as operator valued superdistribution if for at least two $z$-arguments, $z_i $ and $z_j $, we have $x_i \neq x_j $ for $i\neq j $. Let $z_1 ,z_2 ,\ldots ,z_n $ be $n$ points in superspace and suppose that there are $i$ and $j$ with $i\ne j$ such that $x_i \ne x_j $. Then it exists a spacelike plane $E$ in $R^4 $ free of points $x_k $ and such that $x_i $ is in the past and $x_j $ in the future of $E$. Let

\begin{equation}
I=\left \{ i \in  N,x_i \ge E\right \}
\end{equation}
In this case causality gives

\begin{equation}
T_n (z_1 ,z_2 ,\ldots ,z_n )=T_{|I|}(I)T_{|I^c |}(I^c ) 
\end{equation}
The splitting (6.7) does not depend on $E$. This shows that $T_n (z_1 ,z_2 ,\ldots ,z_n )$ is well defined on $R^{4n}\setminus D_n $ where $D_n $ is the diagonal set. The problem now is to extend $T_n $ as superdistribution to all points in $R^{4n}$ (i.e. to extend its components). Due to translation invariance this will be equivalent to an extension at the origin in difference variables. \\
Using (6.7) the Wick theorem provides us with an explicit formula for $T_n $ on $R^{4n}\setminus D_n $ denoted by $\hat T_n $. It is

\begin{gather}
\hat T_n (z_1 ,z_2 ,\ldots ,z_n )=\sum_{l_i <k}\hat t_n^{l_1 ,l_2 ,\ldots ,l_n } (z_1 ,\ldots ,z_n ):\Phi^{k-l_1  }(z_1 ),\ldots ,\Phi^{k-l_n }(z_n ):
\end{gather}
where $\hat t_n^{l_1 ,l_2 ,\ldots ,l_n }(z_1, z_2 ,\ldots ,z_n )$ defined on $ R^{4n}\setminus D_n $ is a scalar superdistribution given by

\begin{gather}
\hat t_n^{l_1 ,l_2 ,\ldots ,l_n }(z_1, z_2 ,\ldots ,z_n )=\sum_{l_{ij}}C(l_{ij})t_{|I|}^{l'_i ,i\in I} 
(I)t_{|I^c|}^{l'_j ,j\in I^c }(I^c )\prod_{i\in I,j\in I^c } \omega (z_i ,z_j )^{l_{ij}}
\end{gather}  
where $C(l_{ij})$ are combinatoric constants and

\begin{gather}
l'_i +\sum_{j\in I^c }l_{ij}=l_i,i\in I \\ 
l'_j +\sum_{i\in I}l_{ij}=l_j,j\in I^c 
\end{gather}
The meaning of $l'_i ,l' _j ,l_{ij} $ is clear. \\
This is the main induction  formula. The two point function $\omega (z)$ is not translation invariant in $z$ but fortunately it is translation invariant in $x$ such that we do not need more powerful tools like wave fronts and even microlocal analysis \cite{BF,F}. The extension of the main formula to the diagonal enables the computation of the scaling degree of $t_n $ for a given $n$ knowing the scaling degree for $n'<n$. It is technically easier to apply the space-time translation invariance in order to reduce the extension problem from the diagonal (in several variables) to an extension problem at the origin. First note that because of traslation invariance in the space-time variables it exist $\hat \tau_n $ with coefficient functions in $\ D'(R^{4n-4}\setminus {0})$ such that

\begin{gather}
\hat \tau_n^{l_1 ,\ldots ,l_n } (x_1-x_2 ,\ldots ,x_{n-1}-x_n ;\theta_1 ,\bar \theta_1 ,\ldots ,\theta_n ,\bar \theta_n )=\hat t_n^{l_1 ,\ldots ,l_n }(z_1 ,\ldots ,z_n )
\end{gather}
The Grassmann variables remain unchanged. We denote $y_i =x_i -x_{i+1},i=1,2,\ldots ,n-1 $ and write for definiteness $I=\left \{1,\ldots ,m \right \},I^c =\left \{m+1,\ldots ,n \right \}$. We get

\begin{gather}\nonumber
\hat \tau_n^{l_1 ,\ldots ,l_n } (y_1 ,\ldots ,y_n ;\theta_1 ,\bar \theta_1 ,\ldots ,\theta_n ,\bar \theta_n )= \\\nonumber
=\sum_{l_{ij}}C(l_{ij})\hat \tau_m^{l'_1 ,\ldots ,l'_m }(y_1 ,\ldots ,y_{m-1} ;\theta_1 ,\bar \theta_1 ,\ldots ,\theta_m ,\bar \theta_m ) \times \\ \nonumber
\times \hat \tau_{n-m}^{l'_{m+1} ,\ldots ,l'_n }(y_{m+1} ,\ldots ,y_{n-1} ;\theta_{m+1} ,\bar \theta_{m+1} ,\ldots ,\theta_n ,\bar \theta_n ) \times \\
 \times \prod _{1\le i\le m, m+1\le j\le n-1}
 \omega^{l_{ij}}(\sum_{r=i}^{j-1} y_r ;\theta_i ,\bar \theta_i ,\theta_j ,\bar \theta_j )
\end{gather}
We follow now \cite{EG,P} undertaking an induction on the scaling degree itself. The idea is to use tensoriality in (6.13) in order to compute the scaling degree at order $n$ from scaling degrees at order $n'<n$ supplemented by scaling degrees of the two point functions. Smearing the two point functions in the $y_m $- variable produces a regular function as in \cite{P} and the scaling in the two point function itself is easy to perform \cite{P}. We came at this stage to a point where we have to interrupt for the moment the flow of analogies with the non-supersymmetric situation because carrying with us nonintegrated $\theta ,\bar \theta $ variables seems at the first glance to make the argument questionable. Indeed there is no translation invariance in 
$\theta ,\bar \theta $ and therefore going to difference variables the decomposition of the time-ordered product of order $n$ in time-ordered products at order $n'<n$ and two point function factors seems to be no longer tensorial. This would make the induction more difficult. But this is not the case because tensoriality is missing only in the Grassmann variables and they behave much more gentle with respect to the scaling as functions of space-time variable do (see also Section  5). One has to perform only an overall counting of them which turns out to be particularly simple because of homogeneity mentioned above. \\
Because the field $\Phi $ can be either $\phi $ or $\bar \phi $ there is a difference as compared to the classical $\phi^4 $-theory which is the standard example of Epstein-Glaser method. This fact has the consequence that the full time ordered products are summs of time-ordered products of third powers of the chiral and antichiral fields (corrected as in (3.23) because of the missing Grassmann integration). We look for such contributions which show largest (supersymmetric) scaling degree. It is clear that in the process of induction they will be associated with two-point functions between $I$ and $I^c $ of maximal scaling degree equal to four. Assume that the scaling degree of $\hat \tau_{n'}^{l_1',l_2',\ldots ,l'_{n'}}$ is smaller or equal to $\sum_{i=1}^{n'} l_i' $. This is verified at $n=2$ because the scaling degree of the two-point function is bounded by four and we have to count in $\phi^3 ,\bar \phi^3 $ missing Grassmann integrations. Then the scaling degree of $\hat \tau_n $ is bounded above by 
\[ \sum_{i=1}^m l'_i +\sum_{i=m+1}^n l'_i +2\sum_{1\le i\le m,m+1\le j\le n-1} l_{ij}=\sum_{i=1}^n l_i\]
Here we corrected again the maximal singularity of the two point function from four to two as above. There is then an extension of $\hat \tau $ to $\tau $ and we have obtained the formula (6.5) with

\begin{gather}
\tau_n^{l_1 ,\ldots ,l_n } (x_1-x_2 ,\ldots ,x_{n-1}-x_n ;\theta_1 ,\bar \theta_1 ,\ldots ,\theta_n ,\bar \theta_n )=t_n^{l_1 ,\ldots ,l_n }(z_1 ,\ldots ,z_n )
\end{gather}
This is the end of the induction. \\
Now we can estimate $\sum_{i=1}^n l_i \leq 2.n $ where $n$ is the order of $T_n $ and the factor 2=3-1 is the corrected third power of the interaction. We substract $4(n-1)-2n$ for the $y,\theta ,\bar \theta $-integrations and get the overall balance of the singular order: $2n-2n+4$. This shows supersymmetric renormalizability for $T_n (x_1 ,\theta _1 ,\bar \theta _1 ;x_2 ,\theta _2 ,\bar \theta _2 ;\ldots ;x_n ,\theta _n ,\bar \theta _n )$. In order to obtain usual renormalizability of the homogeneous integrated 
\begin{gather} \nonumber
T_n (x_1 ,x_2 ,\ldots ,x_n )= \\ \nonumber
=\int d^2 \theta_1 d^2 \bar \theta  \ldots d^2 \theta_n d^2\bar \theta_n T_n (x_1 ,\theta_1 ,\bar \theta_1 ;x_2 ,\theta_2 ,\bar \theta_2 ;\ldots ;x_n ,\theta_n ,\bar \theta_n )
\end{gather}
we substract/add consistently the contribution of the Grassmann
variables and integrations to obtain the upper bound of the singular
order again: $ 2n-2n+4 $. Note that the homogeneity is central for going
from $T_n (z_1 ,z_2 ,\ldots ,z_n )$ to the integrated $T_n (x_1 ,x_2
,\ldots ,x_n )$. From physical work \cite{WB,G,W} we know that this
result is not optimal; the reason is our generosity in managing the Grassmann variables in the process of induction. Nevertheless the bound we have obtained on singular order of the time ordered products means already that the model is renormalizable but not superrenormalizable. This completes our illustration of the causal perturbation theory in the case of the supersymmetric Wess-Zumino model. \\ 
A closer look at the induction for particular time ordered products
should give the exact balance $2n-2n$. The zero singular order is the
indication that we have no mass and no coupling constant renormalization. The only divergences are of logarithmic type and indicate wave function renormalization. The results are consistent with those obtained by studying Feyman supergraphs \cite{WB}. It is belived that the method works also for the abelian case of gauge theories. In this case the transversal sector in our Hilbert-Krein structure has to be substracted like in \cite{FC} in order to impose positive definiteness. This space will be the appropriate scenario of the perturbative (abelian) supersymmetric gauge theory.

\end{document}